\documentstyle[12pt]{article}

\def\be{\begin{equation}}
\def\ee{\end{equation}}
\def\ba{\begin{array}{c}}
\def\baa{\begin{array}{ll}}
\def\ea{\end{array}}

\def\ben{$$}
\def\een{$$}
 
\begin{document}

\titlepage

  \begin{center}{\Large \bf
Matching method and exact solvability of discrete ${\cal
PT}-$symmetric square wells
 }\end{center}

\vspace{5mm}

  \begin{center}

Miloslav Znojil\footnote{ e-mail: znojil@ujf.cas.cz}

 \vspace{3mm}

\'{U}stav jadern\'e fyziky AV \v{C}R, 250 68 \v{R}e\v{z}, Czech
Republic\\

 \vspace{3mm}

\end{center}

\vspace{5mm}


\section*{Abstract}

Discrete ${\cal PT}-$symmetric square wells are studied. Their
wave functions are found proportional to classical Tshebyshev
polynomials {\em of complex argument}. The compact secular
equations for energies are derived giving {\em the real spectra}
in certain intervals of non-Hermiticity strengths $Z$. It is
amusing to notice that although the known square well re-emerges
in the usual continuum limit, a twice as rich, upside-down
symmetric spectrum is exhibited by all its present discretized
predecessors.

 \vspace{9mm}

\noindent PACS  03.65.Ge, 03.65.Ca, 02.30.Tb, 02.30.Hq

\noindent MSC 2000: 81Q05, 81Q10, 46C20, 47B50, 34L40

\vspace{9mm}

  \begin{center}
\end{center}

 \newpage

\section{Introduction}

Undoubtedly, solvable models represent one of the most important
and inspiring sources of insight in the physical properties of
real quantum systems \cite{Fluegge}. For illustration, one may
recollect how the concept of the so called shape-invariant
analytic potentials $V(x)$ \cite{SI} opened the path towards a
better understanding of supersymmetry \cite{Khare}. Recently, a
similar major progress in our understanding of some less standard
structures admitted by the standard Quantum Mechanics \cite{Geyer}
has been initiated by a few numerical as well as non-numerical
analyses \cite{DB,BG,BB} of certain complex potentials exhibiting
another interesting invariance property,
 \be
 V(x) = \left [ V(-x)
 \right ]^*\,.
 \label{SY}
 \ee
It has been argued \cite{BB} that the latter invariance may prove
highly relevant in physics as it makes all the Hamiltonian ${\cal
PT}-$symmetric, i.e., commuting with the product of parity ${\cal
P}$ and complex conjugation ${\cal T}$ while, in its turn, the
latter operator represents the time-reversal operation in the
language of phenomenological considerations.

In the resulting slightly innovated formalism of quantization
(called, for definiteness, ${\cal PT}-$symmetric Quantum
Mechanics, PTSQM \cite{BBjmp,BBJ}), you admit that Hamiltonians
need not necessarily be Hermitian in the usual trivial metric $I$
in Hilbert space,  $H \neq H^\dagger$. As an illustrative example
one may recollect the ordinary differential Schr\"{o}dinger
equation on a finite interval,
 \be
 \left [ -\frac{d^2}{dx^2} + V(x)
 \right ]\,
 \psi(x)=E\,\psi(x), \ \ \ \ \ \ \ \ \
 \psi(\pm 1)=0
 \label{basic}
 \ee
where PTSQM admits potentials $V(x)$ of eq. (\ref{SY}) which are
not real but merely ``parity plus time-reversal" -- symmetric.

One of the most immediate consequences of the invariance
(\ref{SY}) replacing Hermiticity is that the corresponding
Hamiltonian $H \neq H^\dagger$ admits the existence of complex
eigenvalues $E_n \neq E_n^*$. In the language of experimental
physics this simply means a potential {\em instability} of the
whole system, a phenomenon truly encountered, e.g., in
relativistic Quantum Mechanics \cite{relati} or in certain
magnetohydrodynamical systems \cite{Steff}. In the former case the
instability may either mean a complete collapse of the system or a
mere controlled creation of the particle-antiparticle pairs
\cite{relatibe}. An even less catastrophic scenario is encountered
in experimental magnetohydrodynamics where the transitions to the
unstable regime prove reversible as a rule \cite{MHD}.

The potential theoretical consequences of ${\cal PT}-$symmetry
seem equally appealing at present. One of the reasons is that in
eq. (\ref{basic}) the coordinates $x$ themselves may be {\em
complex} and, hence, {\em manifestly} unobservable! This means
that one gets quite close to the relativistic field theory where
the corresponding analogue of $x$ (viz., the field amplitude
$\varphi$) is {\em also} an {\em auxiliary} (i.e., in principle,
not necessarily observable) quantity \cite{BB,BBF}.

In a way inspired by the latter comment the {\em auxiliary} role
of the {\em complex}, unobservable coordinate $x$ may equally well
be played by any other element $x_k$ of any other  unobservable
set ${\cal D} \neq (-\infty,\infty)$. Without getting too deeply
involved in this theoretical question we decided to parallel the
most popular strategy (where ${\cal D}$ is being specified as a
suitable complex curve \cite{BG,BB}) and to try to treat the
(admissible) replacement $ (-\infty,\infty) \to {\cal D}$ as a
{\em simplification}, {\em discretization} of the real axis.

Our present paper summarizes a few of our most interesting
results. Firstly, we set the scene in section \ref{RKs} where we
list some supplementary motivations and show how we intend to
construct our illustrative examples. In the subsequent section
\ref{methody} both the physical and mathematical background of our
considerations is outlined in some necessary detail.

Up to the final brief summary of our results in section
\ref{sumar}, the rest of our paper becomes more technical and it
shows how our simplest discrete models can be solved and how their
various generalizations can be treated and remain, in a current
vague sense of the word, exactly solvable.

\section{PT-symmetry in Quantum Mechanics \label{RKs} }

\subsection{Non-analytic Schr\"{o}dinger equations}

The admissibility of a complexification of spectra is a puzzling
possibility which inspired qualitative analyses of many concrete
models. For many analytic potentials $V(x)$ of an immediate
physical interest the constructive demonstrations of the stability
(i.e., of the reality of the spectrum) have been performed by
perturbative \cite{Rafa}, quasi-classical \cite{Alvarez} as well
as purely numerical \cite{RafaII} methods.

At an early stage of development the {\em non-analytic} solvable
models have been considered conspicuous. Many of them --
typically, potential $V(x) \sim |x|$ on the whole real line,
solvable in terms of Airy functions by the matching method --
happened to exhibit instabilities (i.e., complex energies) at {\em
all} the non-vanishing values of the corresponding coupling
constants \cite{BBjmp}. This caused a certain delay of return to
some of the most elementary exactly solvable piecewise-constant
${\cal PT}-$symmetric potentials in eq.~(\ref{basic}). Still,
their later studies \cite{sqw,Kurasov} clarified that their
spectrum of bound states may remain real in a fairly large domain
of their coupling constants (cf. also \cite{sgezou,myctyri}). In
ref.~\cite{Langer} it has been proved that our ${\cal
PT}-$symmetric Schr\"{o}dinger eq.~(\ref{basic}) possesses {\em
solely} the discrete spectrum of {\em real} energies $E=E_n$,
$n=0, 1, \ldots$ for {\em all} the {complex} potentials which are
``not too strong",
 \be
 \|V\|_\infty <\frac{3}{8}\,\pi^2 \approx 3.701\,.
 \label{potcy}
 \ee
At the same time, the possible emergence of the instabilities {\em
at all the non-zero couplings} has been confirmed for certain
piecewise-constant ${\cal PT}-$symmetric potentials acting on the
complex, curved contours \cite{fragile}.

\subsection{Runge-Kutta option}

The indefiniteness of conclusions of the above observations
inspired out present extension of the study of the
piecewise-constant potentials towards the models which are defined
over a mere {\em discrete}, equidistant lattice of points
 \be
 x_0=-1, \ \ \ \ \ \ \ \ x_k=x_{k-1}+h=-1+kh, \ \ \ \ \ \ \ \
  h = \frac{2}{N}, \ \ \ \ \ \ \ \
  k = 1, 2, \ldots, N\,.
  \label{RK}
 \ee
These models possessing ${\cal PT}-$symmetry and real spectra will
be characterized by the {\em variability} of the number $N+1$ of
grid points in (\ref{RK}). We believe that in this perspective our
understanding of the generic ${\cal PT}-$symmetric models may
acquire a new dimension, returning us to the usual
differential-equation framework in the $N \to \infty$ limit.

One of the most natural possibilities of the discretization of the
differential eq.~(\ref{basic}) is given by the well known
Runge-Kutta recipe \cite{thatwork}
 \be
  -
 \frac{\psi(x_{k+1})-2\,\psi(x_{k})+\psi(x_{k-1})}{h^2}
  +V(x_k)
 \,\psi(x_{k})
 =E\,\psi(x_k)\,.
 \label{diskr}
 \ee
Using the standard boundary conditions
 \ben
  \psi(x_0)=\psi(x_N)=0
  \een
we shall employ here just the piecewise constant, purely imaginary
antisymmetric potentials of refs. \cite{sqw,myctyri},
 \be
 V(x) =
 \left \{
 \begin{array}{ll}
 +{\rm i}\,Z_n & x\in (-\ell_n, -\ell_{n-1} ),
 \\[0.1cm]
 -{\rm i}\,Z_n & x\in (\ell_{n-1},\ell_n ),
 \ea
 \right ., \ \ \ \ \ n = 1, 2, \ldots, q+1\,.
 \label{potenc}
 \ee
Their discontinuities lie at the matching points
$\ell_0\,=0<\ell_1<\ldots<\ell_{q+1}=1$ so that at $q=0$ we have
to solve the discrete Schr\"{o}dinger bound-state problem
 \be
  -
 \frac{\psi(x_{k+1})-2\,\psi(x_{k})+\psi(x_{k-1})}{h^2}
  -{\rm i}\,{\rm sign}(x_k)
 Z\,\psi(x_{k})
 =E\,\psi(x_k),\ \ \ \ \ \ \ \ q=0
 \label{disktrr}
 \ee
etc. It is worth noticing that the first nontrivial $N=4$ version
of eq.~(\ref{disktrr}) coincides  with the Weigert's matrix model
 \be
 \left (
 \begin{array}{ccc}
 2+\frac{1}{4}\,{\rm i}\,{Z}&-1&0\\
 -1&\ \,2&-1\\
 0&-1&2-\frac{1}{4}\,{\rm i}\,{Z}
 \ea
 \right )\,
 \left (
 \ba
 \alpha_0\\
  \gamma\\
 \beta_0
 \ea
 \right )
 =
 \frac{1}{4}\,E
 \,
 \left (
 \ba
 \alpha_0\\
  \gamma\\
 \beta_0
 \ea
 \right )
 \label{matr4}
 \ee
studied in ref.~\cite{Weigert}, in more detail, as ``the simplest
nontrivial" Runge-Kutta discretization of the ${\cal
PT}-$symmetric square well  of ref.~\cite{sqw}.

\section{Physics and mathematics behind the
discrete  Schr\"{o}dinger equations \label{methody} }

\subsection{The definition of the physical metric }

In the light of reviews \cite{Geyer,BBJ}, the ${\cal
PT}-$symmetric Hamiltonians are not compatible with the standard
``Dirac's" metric $I$ in Hilbert space (indeed, we have $H \neq
H^\dagger$). A construction of an appropriate (i.e., of a
Hamiltonian-dependent) generalization $\Theta \neq I$ of the
metric is necessary for a correct probabilistic interpretation of
the measurements. This means that our Hamiltonian only must be
Hermitian (or, more precisely, self-adjoint) with respect to the
generalized metric,
 \be
 H^\dagger = \Theta\,H\,\Theta^{-1}, \ \ \ \ \ \ \ I \neq \Theta
 = \Theta^\dagger >0.
 \label{quasiher}
 \ee
All the physical contents of our models must be derived from the
{new} metric of course. For this reason it makes sense to avoid
confusion by calling eq.~(\ref{quasiher}) with  $\Theta \neq I$ a
``quasi-Hermiticity" condition~\cite{Geyer}.

One can notice that almost exclusively, standard textbooks train
our ``quantum intuition" on the representations of the Hilbert
space where the metric remains trivial. This means that all the
nonstandard PTSQM considerations profit enormously from any
construction of a  sufficiently flexible class of solvable models.
In such a setting our present finite-dimensional ${\cal
PT}-$symmetric models may play the role of a guide towards the
interpretation, especially due to their facilitated connection
with the continuous $N\to \infty$ limit.

A parallel reason for the attention paid here to the discretized
systems lies in the well known difficulty of the problem of the
efficient construction of $\Theta \neq I$ for a given $H\neq
H^\dagger$. Using the Dirac-inspired ``brabraket" compact notation
of ref. \cite{cc} and employing the biorthogonal basis composed of
the left and right eigenstates of our non-Hermitian Hamiltonian
$H$ we may recollect (cf., e.g., \cite{alimet}) that the knowledge
of the spectral representation of the Hamiltonian,
 \ben
 H = \sum_n\,|n\rangle\,\frac{E_n}{\langle\langle n|n\rangle}
 \,\langle\langle n|
 \een
almost immediately leads to the parallel multiparametric formula
 \be
 \Theta = \sum_n\, |n\rangle\rangle\,{\theta_n}
 \,\langle\langle n|\,, \ \ \ \ \ \ \ \ \theta_n>0
 \label{useful}
 \ee
which specifies the correct metric as a (non-unique,
parameter-sequence-dependent) solution of the operator
eq.~(\ref{quasiher}). Thus, we can employ this formula and
construct $\Theta$, more or less comfortably, for the majority of
the exactly solvable models of the form (\ref{basic})
\cite{Batal}. This is in contrast with the situation where one
must construct the states $|n\rangle\rangle$ by some approximative
method. This would make formula (\ref{useful}) practically
useless. Complicated computations would be needed which,
typically, search for $\Theta$ in the form of a product where the
parity ${\cal P}$ is multiplied by a charge ${\cal C}$ (from the
left, \cite{BBJ}) or by a quasi-parity~${\cal Q}$ (from the right,
\cite{pseudo}).

\subsection{Matching method}

Let us recollect that the $N = \infty$ continuous-limit models
(\ref{potenc}) are exactly solvable at any $q\geq 0$ and that
their practical solution remains feasible and transparent at
$q=0$~\cite{sqw}, at $q=1$~\cite{dve} and at $q=3$~\cite{tri} at
least. The solvability in closed form also characterizes the
parallel modifications of these models characterized by the
periodic boundary conditions~\cite{triper}.

It would be highly desirable to extend the available exact
solution of the discrete Weigert's ``minimal" finite-dimensional
model~(\ref{matr4}) with $q=0$ and $N=4$ to some higher integers
$N$ and/or $q$. A key purpose of our present paper is to show that
such extensions are feasible, indeed, once we adapt the matching
method, so efficient in its standard differential-equation form,
to the needs of difference equations.

In the next section our constructive considerations will start
from one of the most straightforward $N>4$ generalizations
 \be
 \left (
 \begin{array}{cccc|c|ccc}
 {\rm i}\xi-F&-1&&&&&&\\
 -1&{\rm i}\xi-F&\ddots&&&&&\\
 &\ddots&\ddots&-1&&&&\\
 &&-1&{\rm i}\xi-F&-1&&&\\
 \hline
 &&&-1&-F&-1&&\\
 \hline
 &&&&-1&-{\rm i}\xi-F&\ddots&\\
 &&&&&\ddots&\ddots&-1\\
 &&&&&&-1&-{\rm i}\xi-F
 \ea
 \right )\,
 \left (
 \ba
 \alpha_0\\
 \alpha_1\\
 \vdots\\
 \alpha_n\\
 \hline
 \gamma\\
 \hline
 \beta_n\\
 \vdots\\
 \beta_0
 \ea
 \right )
 =0
 \label{matr}
 \ee
of the Weigert's model represented by the $N-1=2n+3$ dimensional
matrix transcription of our difference Schr\"{o}dinger
eq.~(\ref{disktrr}) with the re-scaled energy eigenvalues
$F=E\,h^2-2$ and with the re-scaled strength $\xi=Z\,h^2$ of the
non-Hermiticity.

We shall see that such a matrix re-formulation of the matching
recipe proves solvable in closed form at all $n=0, 1, \ldots$. In
the Hermitian context of Quantum Chemistry we just arrive at the
well known H\"{u}ckel's solvable models in the limit $Z \to 0$
\cite{Hueckel}.

\section{Finite-dimensional Schr\"{o}dinger equation~(\ref{matr})
  \label{above}}

\subsection{Closed solvability}

Recollecting the definition of the classical Tshebyshev
polynomials of the second kind \cite{Ryzhik},
 \ben
 U_k(\cos \theta) = \frac{\sin (k+1)\theta}{\sin \theta},
 \ \ \ \ \ \ k = 0, 1, \ldots,
 \een
(cf. a collection of their most relevant properties in Appendix A)
and using the {\em real\,}  elements $\gamma=\psi(0)$, $a_k= {\rm
Re}\,\psi\left ( x_{k+1}\right )$ and $b_k= {\rm Im}\,\psi\left (
x_{k+1}\right )$ we are now prepared to formulate our first
result.

\vspace{.1cm}

 \noindent
{\large \bf Theorem 1}.

 \noindent
Whenever the ${\cal PT}-$symmetry remains unbroken, closed
solutions of eq. (\ref{matr}) are defined by the formulae
 \ben
 \alpha_k=a_k+{\rm i}\,b_k, \ \ \ \ \ \ \
 \beta_k=a_k-{\rm i}\,b_k\equiv \alpha_k^*,
 \een
 \be
 \alpha_k
 = (
 a+{\rm i}b )\,
 U_k
 \left (
 \frac{-F+{\rm i}\xi}{2}
 \right )\,,
 \ \ \ \ \ \ \ \ \ k=0,1,\ldots, n\,
 \label{eigenvecs}
 \ee
and
 \be
 \gamma= \left (
 a_{}+{\rm i}b_{}\right )\,U_{n+1}
 \left (
 \frac{-F+{\rm i}\xi}{2}
 \right )
 =
 \left (
 a_{}-{\rm i}b_{}\right )\,
 U_{n+1}
 \left (
 \frac{-F-{\rm i}\xi}{2}
 \right )
 \label{redmatrbe}
 \ee
complemented by the selfconsistency constraint
 \be
 F\,\gamma= -
 \left (
 a_{}+{\rm i}b_{}\right )\,U_{n}
 \left (
 \frac{-F+{\rm i}\xi}{2}
 \right )
 -
 \left (
 a_{}-{\rm i}b_{}\right )\,
 U_{n}
 \left (
 \frac{-F-{\rm i}\xi}{2}
 \right )\,.
 \label{redmatral}
 \ee

 \noindent
{\large \bf Proof}.

 \noindent
In the first step of our analysis we get the relation $\beta_k=
\alpha_k^*$ due to the ${\cal PT}-$symmetry of our eigenvectors.
We then identify eq. (\ref{matr}) with the three-term recurrences
satisfied by the Tshebyshev polynomials (cf. Appendix A) and
verify formula (\ref{eigenvecs}) for eigenvectors by showing that
it is compatible with the corresponding boundary conditions (at
the smallest subscripts $k$). As long as all this reduces the full
tridiagonal $(N-1) \times (N-1)-$dimensional matrix
eq.~(\ref{matr}) to the mere three matching conditions,
 \be
 \left (
 \begin{array}{ccccc}
 -1&{\rm i}\xi-F&-1&0&0\\
 0&-1&-F&-1&0\\
 0&0&-1&-{\rm i}\xi-F&-1
 \ea
 \right )\,
 \left [
 \ba
 U_{n-1}
 \left (
 \frac{-F+{\rm i}\xi}{2}
 \right )\,
 \left (
 a_{}+{\rm i}b_{}\right )\\
 U_n
 \left (
 \frac{-F+{\rm i}\xi}{2}
 \right )\,
 \left (
 a_{}+{\rm i}b_{}\right )\\
 \gamma\\
 U_{n}
 \left (
 \frac{-F-{\rm i}\xi}{2}
 \right )\,
 \left (
 a_{}-{\rm i}b_{}\right )\\
 U_{n-1}
 \left (
 \frac{-F-{\rm i}\xi}{2}
 \right )\,
 \left (
 a_{}-{\rm i}b_{}\right )
 \ea
 \right ]
 =0
 \label{redmatr}
 \ee
it remains for us to show that the first and the third line may be
simplified to give eq. (\ref{redmatrbe}) while the middle line
defines the product (\ref{redmatral}). QED

\vspace{.1cm}

 \noindent
{\large \bf Remark}.

 \noindent
Without an assumption of an unbroken ${\cal PT}-$symmetry one
would have to admit that the eigenvalues $F$ are complex. A more
explicit illustration of what happens has been described, in the
continuum limit, in ref. \cite{raz}.

\vspace{.1cm}

 \noindent
{\large \bf Corollary}.

 \noindent
A nontrivial solution always exists at $F=0$.

 \noindent
{\large \bf Proof}.

 \noindent
In a discussion of consequence of Theorem 1 one has to notice that
formula  (\ref{redmatral}) forces us to distinguish between two
regimes where $F=0$ and $F \neq 0$, respectively. Once we set,
tentatively, $F=0$, it is easy to deduce from
eq.~(\ref{redmatrbe}) that the parameter $ a_{}$ must vanish for
the even $n = 0, 2, 4, \ldots$ (and we may normalize $b=1$) while
$ b_{}=0$ and $a=1$ for the odd $ n = 1, 3, 5, \ldots$. Thus,
eq.~(\ref{redmatrbe}) degenerates to the mere definition of the
last element $\gamma$ of the eigenvector and we are left with the
single secular eq.~(\ref{redmatral}) which acquires the following
two alternative forms,
 \be
 \ba
 U_{n}
 \left (
 \frac{1}{2}\,{\rm i}\,\xi
 \right )-
 U_{n}
 \left (
 \frac{1}{2}\,{\rm i}\,\xi
 \right )=0,
 \ \ \ \ n = 2m,
 \\
 U_{n}
 \left (
 \frac{1}{2}\,{\rm i}\,\xi
 \right )
 +
 U_{n}
 \left (
-\frac{1}{2}\,{\rm i}\,\xi
 \right )=0,
 \ \ \ \ n = 2m+1.
 \ea
 \label{redmatralbe}
 \ee
These conditions are satisfied identically at any $m = 0, 1,
\ldots$. Thus, our tentative ``guess of the energy" was correct
and that $F=0$ is always an eigenvalue. QED.

\vspace{.1cm}

 \noindent
{\large \bf Remark}.

 \noindent
In spite of the non-Hermiticity of the Hamiltonian, the $F=0$
eigenvalue is ``robust" \cite{dve} and remains real at {\em all}
the real couplings $Z \in (-\infty,\infty)$. The spectrum remains
symmetric with respect to this ``middle point" [note that $F=0$
corresponds to the energy $E=E_{n+2}= 2/h^2=N^2/2=2(n+2)^2$]. The
existence of such a ``central" eigenvalue is not in contradiction
with the differential-equation results of ref. \cite{Langer}.
Indeed, this level moves up with dimension $N$ and disappears in
infinity in the limit $N \to \infty$. We may conclude that in this
sense a ``richer" structure is exhibited by the spectrum at the
finite dimensions.

\subsection{Properties of the ``fragile"  $F \neq 0$ solutions}

Whenever $F \neq 0$ we may treat eq.~(\ref{redmatrbe}) not only as
the condition of vanishing of the imaginary part of $\gamma$,
 \be
  U_{n+1}
 \left (
 \frac{-F+{\rm i}\xi}{2}
 \right )\,
 \left (
 a_{}+{\rm i}b_{}\right )
 =
 U_{n+1}
 \left (
 \frac{-F-{\rm i}\xi}{2}
 \right )\,
 \left (
 a_{}-{\rm i}b_{}\right )
 \label{redmatrbex}
 \ee
but also as an explicit definition of the non-vanishing
left-hand-side quantity $F\gamma$ in eq.~(\ref{redmatral}). Its
insertion simplifies the latter relation,
 \be
 T_{n+1}
 \left (
 \frac{-F+{\rm i}\xi}{2}
 \right )\,
 \left (
 a_{}+{\rm i}b_{}\right )
 =-
 T_{n+1}
 \left (
 \frac{-F-{\rm i}\xi}{2}
 \right )\,
 \left (
 a_{}-{\rm i}b_{}\right )
 \label{redmatrbexbfi}
 \ee
where $T_k(z)$ denotes the $k-$th Tshebyshev polynomial of the
first kind.

One of the latter two relations defines the normalization vector
$(a,b) = (a_0,b_0)$ while their ratio gives
 \be
 T_{n+1}
 \left (
 \frac{-F+{\rm i}\xi}{2}
 \right )\,U_{n+1}
 \left (
 \frac{-F-{\rm i}\xi}{2}
 \right )
+
 T_{n+1}
 \left (
 \frac{-F-{\rm i}\xi}{2}
 \right )\,U_{n+1}
 \left (
 \frac{-F+{\rm i}\xi}{2}
 \right )=0.
 \label{redfinal}
 \ee
This represents our final secular equation which defines, in an
implicit manner, the energies $F$ as functions of the
couplings~$\xi$.

An efficient numerical treatment of the latter eigenvalue problem
may be based on the re-parametrization
 \be
  \frac{-F+{\rm i}\xi}{2} = \cos \varphi, \ \ \ \ \ \ \ \
  {\rm Re}\,\varphi = \alpha, \ \ \ \ \ \ \
  {\rm Im}\,\varphi = \beta
  \label{mapping}
  \ee
i.e.,
 \be
 \frac{1}{2}\,F=-\cos\alpha \cosh\beta,
 \ \ \ \ \ \ \
 \frac{1}{2}\,\xi=-\sin\alpha \sinh\beta
 \label{dvaro}
 \ee
and, in the opposite direction,
 \ben
 \cos\alpha=
 -\frac{1}{2\cosh\beta}\,F ,
 \ \ \ \ \ \ \
 \sinh \beta =\frac{1}{2\sqrt{2}}\,
 \sqrt{F^2+\xi^2-4+\sqrt{(F^2+\xi^2-4)^2+16\,\xi^2}}.
 \een
This change of variables transforms eq.~(\ref{redfinal}) into the
trigonometric secular equation
 \be
  {\rm Re}\,\frac{\sin [(n+1)\varphi]
   \cos [(n+1)\varphi^*]}{\sin \varphi}=0.
   \label{tojevon}
  \ee
Its roots must be determined, in general, numerically.

\vspace{.1cm}

 \noindent
{\large \bf Lemma}.

 \noindent
In the domain with negative $\beta<0$ the roots of eq.
(\ref{tojevon}) have $\alpha \in (0, \pi/2)$ at the negative $F<0$
and  $\alpha \in (\pi/2,\pi)$ at the positive $F>0$.

 \noindent
{\large \bf Proof}.

 \noindent
The constant value of the coupling $\xi>0$ is mapped upon a
downwards-oriented half-oval in the $\alpha-\beta$ plane. Its top
lies at $\alpha=\pi/4$ while its two asymptotes $\alpha= 0$ and
$\alpha = \pi/2$ are reached in the limit $\beta \to -\infty$. In
such a representation the ``robust" and $\xi-$independent energy
level $F=0$ lies on the top of the half-oval while its decreasing
and increasing neighbors are found displaced to the left and
right, respectively, along the half-oval downwards. QED.

\vspace{.1cm}

 \noindent
{\large \bf Remark}.

 \noindent
At the first few lowest dimensions $N-1=2n+3$ the roots of eq.
(\ref{tojevon}) may be written in closed form,
 \ben
 \ba
 F_0=0, \ \ F_\pm =\pm \sqrt{2-\xi^2}, \ \ \ \ n = 0,\\
 F_0=0, \ \ F_{\pm,\pm} =\pm \sqrt{2-\xi^2\pm \sqrt{1-4\xi^2}},
  \ \ \ \ \ \ n = 1
 \ea
 \een
etc. This enables us to determine the respective critical values
$Z=Z_{(crit)}(N)$, i.e., the exact
 $
 Z_{crit}(4)=4\,\sqrt{2} \approx 5.66$ at $ n = 0$ and $
 Z_{crit}(6)=9/{2} =4.50$ at $n = 1$
followed by the numerically calculated values $Z_{crit}(8)=4.463$
at $n=2$, $Z_{crit}(10)=4.461$ at $n=3$  and $Z_{crit}(12)=4.463$
at $n=4$ etc. These results do not contradict the expected $n \to
\infty$ limit as derived in ref.~\cite{raz},
$Z_{crit}(\infty)=4.475$ (cf. Figure 1).

\section{The role of the integers $N$ and $q$ \label{remark} }

\subsection{Closed solvability at the odd $N = 2n+3$ }

In comparison with ref. \cite{Weigert} a ``one-step easier"
discretization of eq.~(\ref{basic}) emerges at $q=0$ and at the
odd $N = 3$, with the two energy roots. A non-equivalent discrete
alternative to eq.~(\ref{matr}) could then read, in the same
notation,
 \be
 \left (
 \begin{array}{cccc|ccc}
 {\rm i}\xi-F&-1&&&&&\\
 -1&{\rm i}\xi-F&\ddots&&&&\\
 &\ddots&\ddots&-1&&&\\
 &&-1&{\rm i}\xi-F&-1&&\\
 \hline
 &&&-1&-{\rm i}\xi-F&\ddots&\\
 &&&&\ddots&\ddots&-1\\
 &&&&&-1&-{\rm i}\xi-F
 \ea
 \right )\,
 \left (
 \ba
 \alpha_0\\
 \alpha_1\\
 \vdots\\
 \alpha_n\\
 \hline
 \alpha^*_n\\
 \vdots\\
 \alpha^*_0
 \ea
 \right )
 =0.
 \label{patr}
 \ee
This leads to the following alternative result.

\vspace{.1cm}

 \noindent
{\large \bf Theorem 2}.

 \noindent
Whenever the ${\cal PT}-$symmetry remains unbroken, closed
solutions of eq. (\ref{patr}) are defined by formulae
(\ref{eigenvecs}) accompanied by the alternative matching
condition
 \be
 \left (
 a_{}+{\rm i}b_{}\right )\,U_{n+1}
 \left (
 \frac{-F+{\rm i}\xi}{2}
 \right )
 =
 \left (
 a_{}-{\rm i}b_{}\right )\,
 U_{n}
 \left (
 \frac{-F-{\rm i}\xi}{2}
 \right )\,.
 \label{redmatrbeno}
 \ee
and by the odd$-N$ counterpart of eq.~(\ref{redfinal}),
 \be
 U_{n}
 \left (
 \frac{-F+{\rm i}\xi}{2}
 \right )\,U_{n}
 \left (
 \frac{-F-{\rm i}\xi}{2}
 \right )
 =U_{n+1}
 \left (
 \frac{-F+{\rm i}\xi}{2}
 \right )\,
 U_{n+1}
 \left (
 \frac{-F-{\rm i}\xi}{2}
 \right )\,.
 \label{redfinalbe}
 \ee

 \noindent
{\large \bf Proof}.

 \noindent
Definition (\ref{eigenvecs}) of the eigenvectors remains unchanged
but the matching condition generated by the subproblem
 \be
 \left (
 \begin{array}{cccc}
 -1&{\rm i}\xi-F&-1&0\\
 0&-1&-{\rm i}\xi-F&-1
 \ea
 \right )\,
 \left [
 \ba
 U_{n-1}
 \left (
 \frac{-F+{\rm i}\xi}{2}
 \right )\,
 \left (
 a_{}+{\rm i}b_{}\right )\\
 U_n
 \left (
 \frac{-F+{\rm i}\xi}{2}
 \right )\,
 \left (
 a_{}+{\rm i}b_{}\right )\\
 U_{n}
 \left (
 \frac{-F-{\rm i}\xi}{2}
 \right )\,
 \left (
 a_{}-{\rm i}b_{}\right )\\
 U_{n-1}
 \left (
 \frac{-F-{\rm i}\xi}{2}
 \right )\,
 \left (
 a_{}-{\rm i}b_{}\right )
 \ea
 \right ]
 =0
 \label{redmatrows}
 \ee
is just one, $\alpha_{n+1}=\alpha_n^*$. The ratio between this
equation and its Hermitian conjugate eliminates all the
normalization ambiguities and leads to the secular equation
(\ref{redfinalbe}). QED.

\vspace{.1cm}

 \noindent
{\large \bf Remark}.

 \noindent
Secular equation (\ref{redfinalbe}) is an implicit definition of
the $N=2n+3$ energy levels $F = F(\xi)$. It possesses the compact
analytic solutions at the first few values of $n$ of course,
 \ben
 \ba
  F_\pm =\pm \sqrt{1-\xi^2}, \ \ \ \ n = 0,\\
  F_{\pm,\pm} =\pm \frac{1}{2}\,\sqrt{6-4\xi^2\pm 2\,\sqrt{5-16\xi^2}},
  \ \ \ \ \ \ n = 1.
 \ea
 \een
The respective elementary expressions for the critical constants
 \ben
 \ba
 Z_{crit}(3)=\frac{9}{4}=2.25, \ \ \ \ n = 0,\\
 Z_{crit}(5)=\frac{25\,\sqrt{5}}{16} \approx 3.49,
  \ \ \ \ \ \ n = 1
 \ea
 \een
are followed by the complex Cardano representation of the real
$Z_{crit}(7)=3.946$ at $n=2$. At $n > 2$ one switches to a purely
numerical algorithm giving $Z_{crit}(9)=4.148$ at $n=3$ etc. In
comparison with the parallel results derived at even~$N$ in
section~\ref{above} above we notice that the numerical convergence
towards  $n = \infty$ is perceivably slower and from the opposite
side at the odd $N$ (cf. Figure~1 once more).

\subsection{Models with more matching points }

New features of the spectrum emerge when we choose $q>0$ in
eq.~(\ref{potenc}). In the light of the differential-equation
results \cite{dve,tri} one may expect the existence of the so
called `robust' levels at $q \geq 1$ and $N < \infty$. By
definition they remain real at the arbitrarily large coupling
constants $Z_k$. {\it A priori}, the position of these levels may
be expected to be controlled by both the parameters $Z_k$ and
$\ell_k$.


Obviously, the interpretation of the discretization becomes
facilitated when we choose the discontinuities at the simple
rational numbers $\ell_k$. It is interesting to note, marginally,
that their choice simplifies even the trigonometric secular
determinants in many differential-equation models. Nevertheless,
in our present context it is much more important that the choice
of the simplest $\ell_k$ at $q \geq 1$ enables us to continue
working with the  not too large matrix dimensions $N-1$.

\subsubsection{$\ell=1/2$ and $N=6$}

For the $q=1$ interaction model (\ref{potenc}) with $\ell_1=1/2$,
 \be
 V(x) =
 \left \{
 \begin{array}{l}
 +{\rm i}\,Z
 \\[0.1cm]
 0
 \\[0.1cm]
 -{\rm i}\,Z
 \ea
 \right .
 \ \ \ \ {\rm for}\ \ \ \
 x \in
  \left \{
 \begin{array}{l}
  \left (-1,-\frac{1}{2} \right ),
 \\[0.1cm]
 \left (-\frac{1}{2},\frac{1}{2} \right ),
 \\[0.1cm]
\left (\frac{1}{2},1 \right )
 \ea
 \right .
 \label{potencalas}
 \ee
we may start from its discretization (\ref{diskr}) with the most
elementary choice of $N=6$,
 \be
 \left (
 \begin{array}{c|ccc|c}
 {\rm i}\xi-F&-1&&&\\
 \hline
 -1&-F&-1&&\\
 &-1&-F&-1&\\
 &&-1&-F&-1\\
 \hline
 &&&-1&-{\rm i}\xi-F
 \ea
 \right )\,
 \left (
 \ba
 \alpha_0\\
 \hline
 \gamma_0\\
 \gamma\\
 \gamma_0^*\\
 \hline
 \alpha^*_0
 \ea
 \right )
 =0.
 \label{matr6}
 \ee
This equation determines the following five eigenvalues,
 \ben
 F_0=0, \ \ \ \
 F_{\pm,\pm}=
 \pm \frac{1}{2}
 \sqrt{8-2\,\xi^2 \pm 2\,\sqrt{4+\xi^4}}.
 \een
Besides the constant and $\xi-$independent level $F_0=0$ one can
show that the two ``external" levels
 \ben
F_{\pm,+} = \pm \sqrt {3} \left [ 1-\frac{1}{12}\,{\xi}^{2}+{\frac
{5}{288}}\,{\xi}^{4}+{ \frac {5}{3456}}\,{\xi}^{6}+O\left
({\xi}^{8}\right ) \right ]
 \een
remain  robustly real for all the values of $\xi$ and never
complexify,
 \ben
F_{\pm,+} = \pm \sqrt {2} \left [
1+\frac{1}{4}\,{y}^{2}-\frac{1}{32}\,{y}^{4}-{\frac {31}{
128}}\,{y}^{6}+O\left ({y}^{8}\right )\right ], \ \ \ \ \ y =
1/\xi\,.
  \een
They are complemented by the fragile pair of levels $F_{\pm,-}$
which coincide with $F_0=0$ at the critical value of $\xi_{crit}=
\sqrt{3/2} \approx 1.2247$ and complexify beyond this
``exceptional" point.

\subsubsection{$\ell=1/2$ and $N=8$ and more}

The $N=8$ calculations lead to the observation that the two
outmost energy levels as well as the ``middle" $F=0$ level remain
robust and real at all the real $\xi$. At the critical value of
$\xi_{crit}= 0.845479352$ we observe the confluence and subsequent
complexification of both the remaining (i.e., positive and
negative) energy pairs at the respective two numerical values of
$F_{crit}=\pm 1.05167218$. With the further growth of $\xi$ the
quantity Re $F^2$ becomes negative  beyond
$\xi_{zero}=3.2222152046$, with the purely imaginary
crossing-point $F_{zero}=\pm 2.146638195\,{\rm i}$.


At $N=10$ the secular algebraic equation for the energy-level
characteristics $y=F^2$ has the following four roots in the
Hermitian limit with $\xi=0$,
 \ben
 y_{a,\pm}=\frac{5}{2} \pm \frac{1}{2}\,\sqrt{5}, \ \ \ \ \ \
 y_{b,\pm}=\frac{3}{2} \pm \frac{1}{2}\,\sqrt{5}.
 \een
Numerically one finds that the real quadruplet $$0.5173571919,
1.810807242, 1.810964520, 3.356678112$$
 of these
values gives the single roots
 $F=\pm 0.7192754632$ and $F=\pm 1.832123935$ and the double roots
            $F=\pm 1.345662380$ at $\xi_{subcrit}=0.50209209$.
            In parallel, the complex
quadruplet $$0.5173571921, 1.810885878 \pm 0.00005268889724\,{\rm
i},
 3.356678117$$
emerges numerically at the upper estimate of the critical coupling
strength $\xi_{supercrit}=0.502092091$.

Similar observations have been made also at $N=12$ and $N=14$ and
all of them parallel and reproduce the qualitative features as
observed in ref.~\cite{dve} in the $N \to \infty$ continuum limit.


\section{The role of the shift-parameter $\ell$  \label{cermak} }

\subsection{Suppressing the non-Hermiticity using $\ell=5/8$
}

In order to illustrate the emergence of a pair of the robust
levels let us re-consider the $q=1$ interaction model
(\ref{potenc}) and employ a larger shift $\ell_1=5/8$. In the
resulting potential
 \be
 V(x) =
 \left \{
 \begin{array}{l}
 +{\rm i}\,Z
 \\[0.1cm]
 0
 \\[0.1cm]
 -{\rm i}\,Z
 \ea
 \right .
 \ \ \ \ {\rm for}\ \ \ \
 x \in
  \left \{
 \begin{array}{l}
  \left (-1,-\frac{5}{8} \right ),
 \\[0.1cm]
 \left (-\frac{5}{8},\frac{5}{8} \right ),
 \\[0.1cm]
\left (\frac{5}{8},1 \right )
 \ea
 \right .
 \label{potencalbe}
 \ee
the non-Hermiticity is weakened at a fixed coupling $Z$. We get
the matrix analogue of problem (\ref{matr6}) from the discretized
eq.~(\ref{diskr}) with $N=8$,
 \be
 \left (
 \begin{array}{c|ccccc|c}
 {\rm i}\xi-F&-1&&&&&\\
 \hline
 -1&-F&-1&&&&\\
 &-1&-F&-1&&&\\
 &&-1&-F&-1&&\\
 &&&-1&-F&-1&\\
 &&&&-1&-F&-1\\
 \hline
 &&&&&-1&-{\rm i}\xi-F
 \ea
 \right )\,
 \left (
 \ba
 \alpha_0\\
 \hline
 \gamma_1\\
 \gamma_0\\
 \gamma\\
 \gamma_0^*\\
 \gamma_1^*\\
 \hline
 \alpha^*_0
 \ea
 \right )
 =0.
 \label{matrb8}
 \ee
An inspection of the explicit form of the secular determinant of
this equation,
 \be
 {\cal D}=
 \left [
 -F^6
 -F^4
 \left (
 \xi^2-6
 \right )
 +F^2
 \left (
 4\,\xi^2-10
 \right )
 -3\,\xi^2+4
 \right ] F\,,
 \label{figureA}
 \ee
confirms that all of its seven roots remain real in the Hermitian
$\xi \to 0$ limit,
 \be
 F_0=0, \ \ \ \
 F_{\pm,0}=
 \pm \sqrt{2}, \ \ \ \
 F_{\pm,\pm}= \pm
 \sqrt{2 \pm \sqrt{2}}, \ \ \ \ \ \ \ \xi \to 0.
 \label{herie}
 \ee
Moreover, in a way illustrated in Figure 2, only two roots
$F_{\pm,-}$ merge and complexify beyond $\xi_{crit}\approx
1.15470$. In the other words, as many as five of the roots remain
real for $Z\gg 1$,
 \ben
 F_0=0, \ \ \ \
 F_{\pm,0}=
 \pm 1, \ \ \ \
 F_{\pm,+}= \pm
 \sqrt{3}, \ \ \ \ \ \ \ \xi \to \infty.
 \een
This observation parallels the similar results obtained in the
continuous limit. All this also agrees with the {\it a priori}
expectations, reflecting the weakened influence of the
non-Hermitian part of the potential at the comparatively large
shift $\ell=5/8$.

\subsection{Strengthening the non-Hermiticity using $\ell=3/8$
}

The number of the  robust levels may vary with the width $\ell$ of
suppression of the imaginary interaction. Using a weaker
suppression $\ell_1=3/8$ we may modify eq.~(\ref{potencalbe})
accordingly and we get the precise $N=8$ analogue
 \be
 \left (
 \begin{array}{c|ccccc|c}
 {\rm i}\xi-F&-1&&&&&\\
 -1&{\rm i}\xi-F&-1&&&&\\
 \hline
 &-1&-F&-1&&&\\
 &&-1&-F&-1&&\\
 &&&-1&-F&-1&\\
 \hline
 &&&&-1&-{\rm i}\xi-F&-1\\
 &&&&&-1&-{\rm i}\xi-F
 \ea
 \right )\,
 \left (
 \ba
 \alpha_0\\
 \alpha_1\\
 \hline
 \gamma_0\\
 \gamma\\
 \gamma_0^*\\
 \hline
 \alpha_1^*\\
 \alpha^*_0
 \ea
 \right )
 =0
 \label{matrce8}
 \ee
of problem (\ref{matrb8}). It is easy to evaluate the related
secular determinant
 \be
 {\cal D}=
 \left [
 -F^6
 -F^4
 \left (
 2\,\xi^2-6
 \right )
 +F^2
 \left (-\xi^4+
 4\,\xi^2-10
 \right )+2\,\xi^4+\xi^2+4
 \right ] F
 \label{figureB}
 \ee
which gives the same roots (\ref{herie}) in the
weak-non-Hermiticity-limit $\xi \to 0$. A change may be expected
to emerge in the more non-Hermitian regime. Quantitatively this
change is being well illustrated by  Figure 3 where we see that
merely two non-vanishing roots (viz., $F_{\pm,+} \to \pm
\sqrt{2}$) remain robust and real at large $Z \to \infty$. We see
that two complexifications take place at a certain finite coupling
$Z$, i.e., at a unique critical point $\xi_{crit} \approx
0.5875691807$. In contrast to the previous case we witness two
{\em distinct} mergers of the eigenvalues $F_{\pm,0}$ with
$F_{\pm,-}$ at the two different exceptional points
$F^{exc.}_{\pm} \approx \pm 1.140716421$.

\section{Summary \label{sumar} }

A tentative weakening of the standard Hermiticity $H=H^\dagger$ to
the mere ${\cal P}-$pseudo-Hermiticity of the Hamiltonian,
 \be
 H^\dagger = {\cal P}\,H\,{\cal P}^{-1}
 \label{pseudoher}
 \ee
employs only too often the ordinary differential Hamiltonian $H$
{\em and} the parity operator ${\cal P}$ \cite{BB}. Here we
intended to emphasize that it is by far not the only possible
scenario. For this reason we replaced eq.~(\ref{basic}) by its
discrete counterpart (\ref{diskr}) and explained that its use may
bring a few important advantages.

We feel satisfied by the fact that the purely formal application
of the PTSQM recipe survives quite easily our present narrowing of
one's attention to the very specific and {\em solvable}
square-well models. We have shown that the use of various specific
discrete versions of the Hamiltonians leads to very relevant
formal simplifications emphasized throughout the text.

Within the dynamical region where our models possess the real
spectrum, very close parallels were shown to emerge between the
bound states for the standard continuous coordinate
$x$~\cite{sqw,dve,tri} and for its present less standard discrete
counterpart(s).

We have seen that in the discrete case it is particularly easy to
complement the validity of the {\em auxiliary}, technical
pseudo-Hermiticity condition~(\ref{pseudoher}) by the constructive
solution of the {\em vital and essential} quasi-Hermiticity
requirement~(\ref{quasiher}) which is of key relevance in the
context of physics.

\subsection*{Acknowledgment}

Work partially supported from IRP AV0Z10480505 and by M\v{S}MT
\v{C}R, project ``Doppler Institute" Nr. LC06002.

\newpage

\section*{Figure captions}

\subsection*{Figure 1. Numerical values and convergence
of the critical couplings $Z_{(crit)}=Z_{(crit)}(N)$.}

\subsection*{Figure 2. The $\xi-$dependence of the seven roots of
the sample secular determinant (\ref{figureA}) }

\subsection*{Figure 3. The $\xi-$dependence of the seven roots of
the sample secular determinant (\ref{figureB}) }

\newpage

\subsection*{Appendix A: Tshebyshev polynomials of scalars and
matrices }

A ``pedestrian's" way to the simultaneous work with the
Tshebyshev's polynomials $Q_n(x)$ of the first kind [replace
$Q_n(x) \to T_n(x)$] and of the second kind [replace $Q_n(x) \to
U_n(x)$] leads through the ``universal" three-term recurrence
relation
 \be
 Q_{n+1}(x) -2x\,Q_n(x) + Q_{n-1}(x)=0
 \ee
complemented by the specific initial values,
 \ben
 U_0(x)=0, \ \
 U_1(x)=\sqrt{1-x^2}, \ \ \ldots
 \een
 \ben
 T_0(x)=1, \ \ T_1(x)=x, \ \ \ldots\,.
 \een
Many useful identities valid for these polynomials are available
in the literature.

For complex arguments, an interesting situation emerges during a
real-matrix re-arrangement of eq.~(\ref{matr}). We may split
eq.~(\ref{matr}) in its real and imaginary parts and `glue' them
together in a pentadiagonal eigenvalue problem
 \be
 \left (
 \begin{array}{cc|cc|cc|cc|c}
 -F&-\xi&-1&0&&&&&\\
 \xi&-F&0&-1&&&&&\\
 \hline
 -1&0&-F&-\xi&\ddots&&&&\\
 0&-1&\xi&-F&&\ddots&&&\\
 \hline
 &&\ddots&&\ddots&\ddots&-1&0&\\
 &&&\ddots&\ddots&\ddots&0&-1&\\
 \hline
 &&&&-1&0&-F&-\xi&-1\\
 &&&&0&-1&\xi&-F&0\\
 \hline
 &&&&&&-2&0&-F
 \ea
 \right )\,
 \left (
 \ba
 a_0\\
 b_0\\
 \hline
 a_1\\
 b_1\\
 \hline
 \vdots\\
 \vdots \\
 \hline
 a_n\\
 b_n\\
 \hline
 \gamma
 \ea
 \right )
 =0.
 \label{rematr}
 \ee
It can be written in a partitioned form,
 \be
 \left (
 \begin{array}{ccccc}
 {\rm \bf X}&{\rm \bf -1}&&&\\
 {\rm \bf -1}&{\rm \bf X}&\ddots&&\\
 &\ddots&\ddots&{\rm \bf -1}&\\
 &&{\rm \bf -1}&{\rm \bf X}&\vec{\rm \bf d}\\
 &&&2\vec{\rm \bf d}^{\,T}&-F
 \ea
 \right )\,
 \left (
 \ba
 {\rm \bf \vec{c}_0}\\
 {\rm \bf \vec{c}_1}\\
 \vdots\\
 {\rm \bf \vec{c}_n}\\
 \gamma
 \ea
 \right )
 =0,
 \label{pamatr}
 \ee
with an `odd' anomalous row and column containing an auxiliary
two-dimensional vector [$\vec{\rm \bf d}^{\,T}=(1,0)$].

In the new context all our wave functions are again proportional
to the Tshebyshev polynomials,
 \be
 {\rm \bf \vec{c}_k} =U_k
 \left (
 \frac{1}{2}
 {\rm \bf X}
 \right )\,
 {\rm \bf \vec{c}_0}\,,
 \ \ \ \ \ \ \
 {\rm \bf X}=
 \left (
 \begin{array}{cc}
 -F&-\xi\\
 \xi&-F
 \ea
 \right )\,,
 \ \ \ \ \ {\rm \bf k} =0,1,\ldots,n+1\,.
 \label{closedf}
 \ee
The existence of the explicit solutions (\ref{closedf}) [where
polynomials $U_k$ depend on a {\em two-by-two matrix} argument
$X$] reduces eq.~(\ref{pamatr}) to the constraint imposed upon the
two-dimensional real vector ${\rm \bf \vec{c}_{n+1}}$. Of course,
this condition is equivalent to the complex matching as mentioned
above.

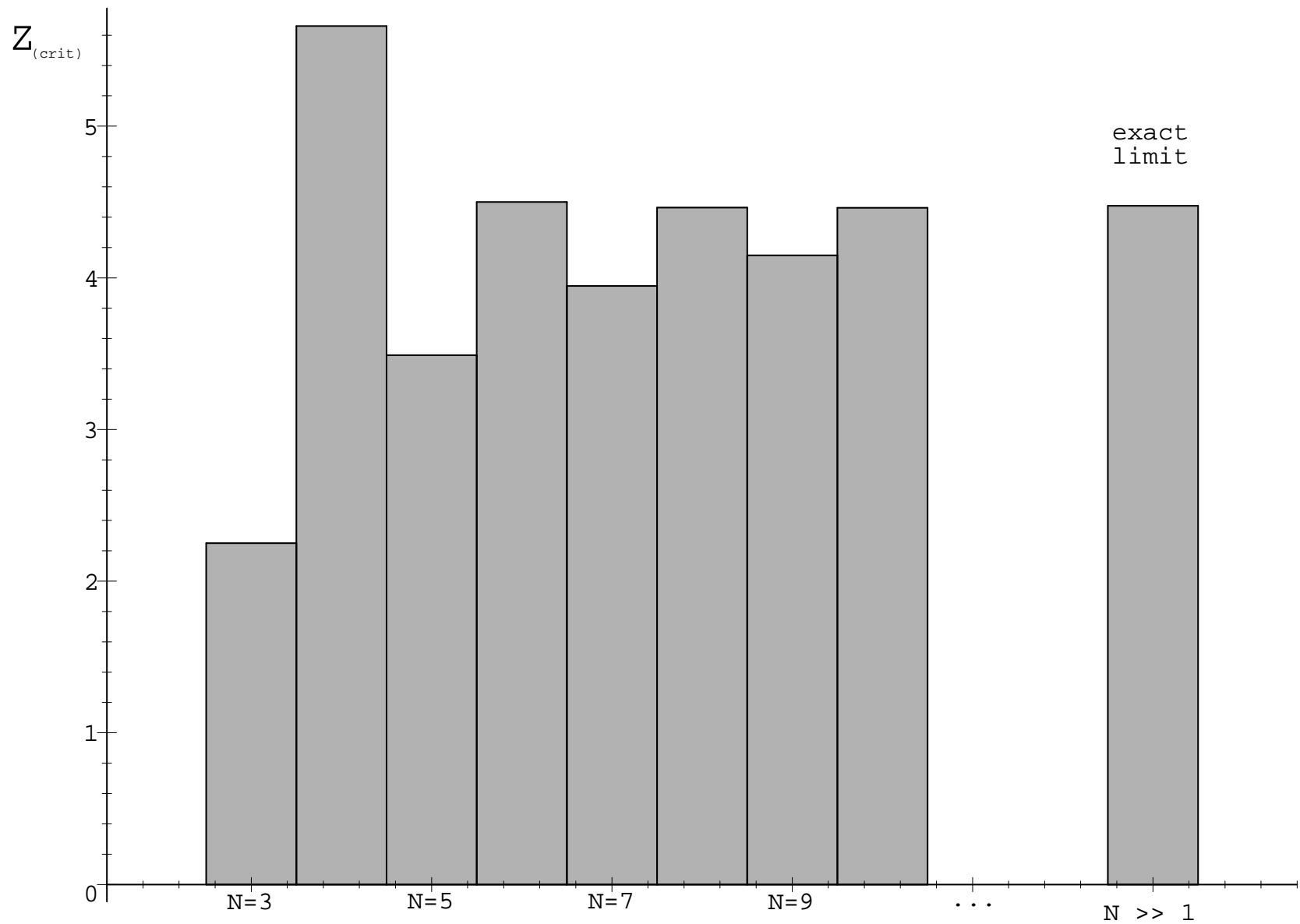
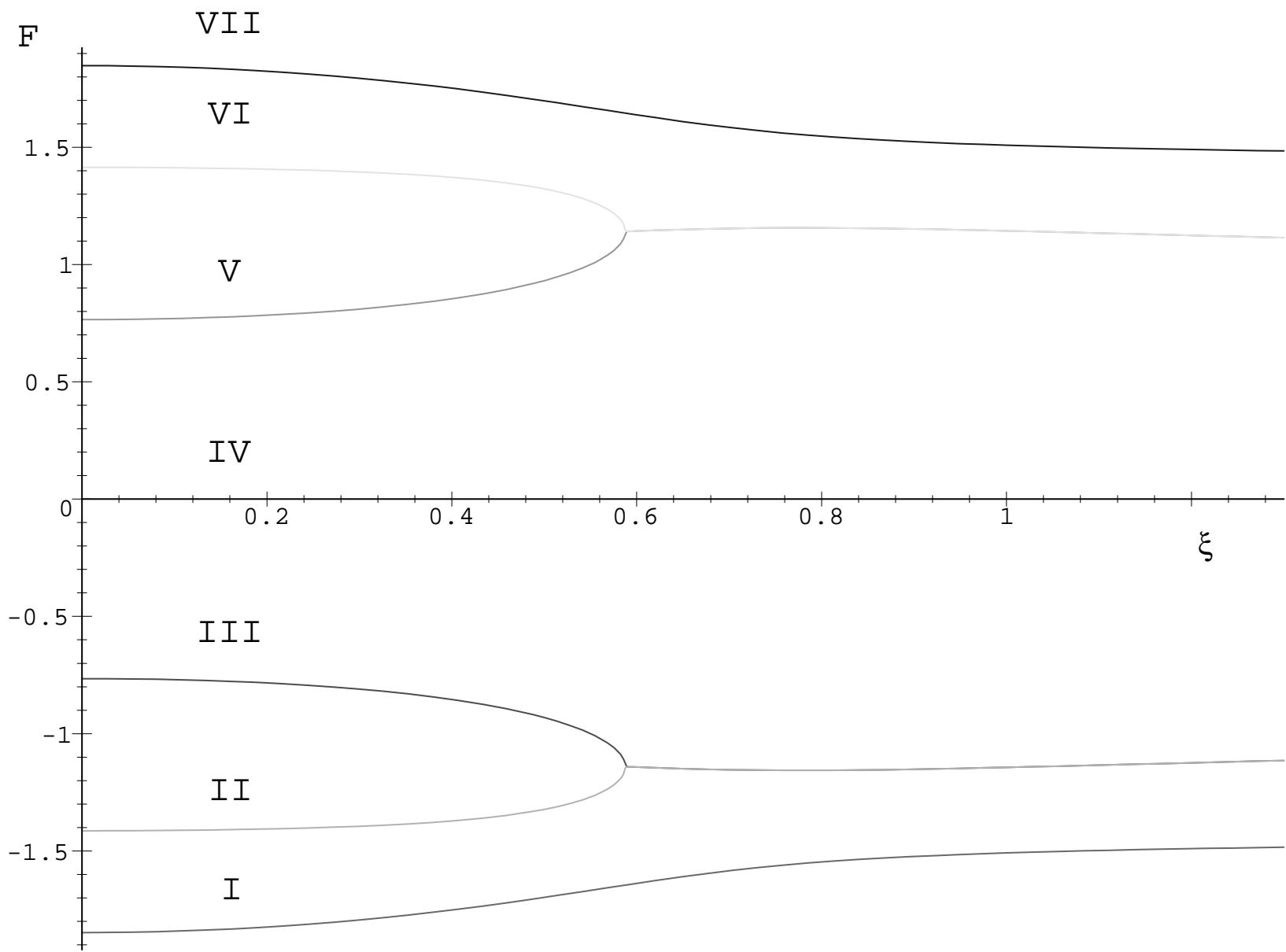
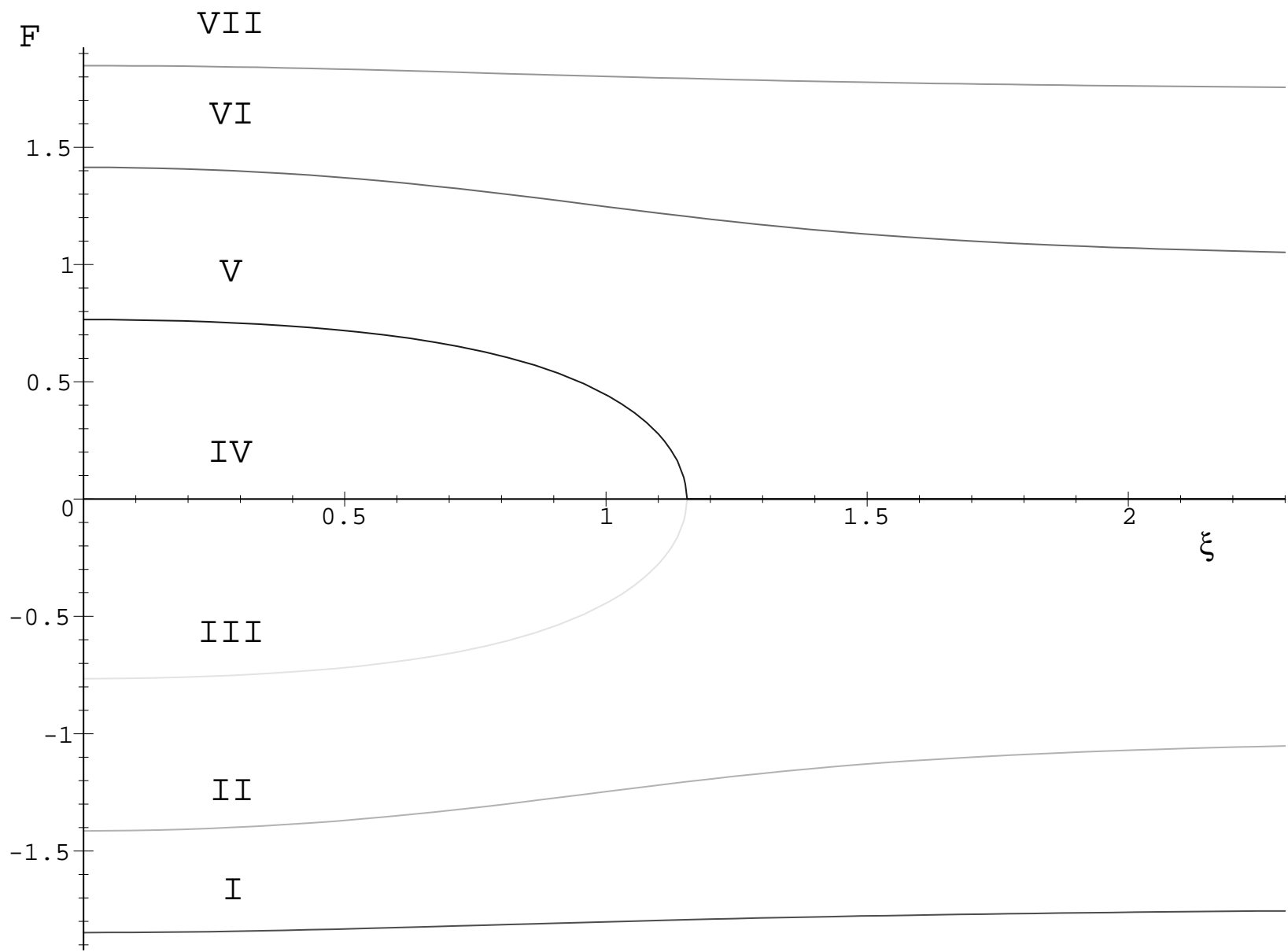
\end{document}